\documentclass[10pt,conference]{IEEEtran}
\IEEEoverridecommandlockouts
\usepackage{cite}
\usepackage{amsmath,amssymb,amsfonts}
\usepackage{graphicx}
\usepackage{textcomp}
\usepackage{xcolor}
\usepackage{algpseudocode}
\usepackage{algorithm,algorithmicx,algcompatible}
\usepackage{braket}
\usepackage{booktabs,multirow}
\usepackage{bm}

\def\BibTeX{{\rm B\kern-.05em{\sc i\kern-.025em b}\kern-.08em
    T\kern-.1667em\lower.7ex\hbox{E}\kern-.125emX}}
\usepackage{adjustbox}
\usepackage{tikz}
\usetikzlibrary{external}

\usepackage[colorlinks=true]{hyperref}

\usepackage[switch]{lineno}
\newcommand*\patchAMSlineno[1]{
                \expandafter\let\csname old#1\expandafter\endcsname\csname #1\endcsname
                \expandafter\let\csname oldend#1\expandafter\endcsname\csname end#1\endcsname
                \renewenvironment{#1}
                                {\linenomath\csname old#1\endcsname}
                                {\csname oldend#1\endcsname\endlinenomath}
}
\AtBeginDocument{
                \patchAMSlineno{equation}
                \patchAMSlineno{equation*}
                \patchAMSlineno{align}
                \patchAMSlineno{align*}
}

\usepackage{algorithm}
\usepackage{graphicx}
\usepackage{float}
\usepackage{braket}
\usepackage{amsmath}
\usepackage{amssymb}
\usepackage{subcaption}
\usepackage{dblfloatfix}
\usepackage[qm]{qcircuit}
\usepackage{graphicx}
\usepackage{tikz}
\usepackage{hyphenat}
\begin{document}
\title{Hierarchical Multigrid Ansatz for Variational Quantum Algorithms}

\author{\IEEEauthorblockN{Christo~Meriwether~Keller\IEEEauthorrefmark{1}\IEEEauthorrefmark{2}, Stephan~Eidenbenz\IEEEauthorrefmark{2}, Andreas~B\"artschi\IEEEauthorrefmark{2},\\ Daniel~O'Malley\IEEEauthorrefmark{3}, John~Golden\IEEEauthorrefmark{2}, Satyajayant Misra\IEEEauthorrefmark{1}}

 \IEEEauthorblockA{\IEEEauthorrefmark{1}Department of Computer Science, New Mexico State University, Las Cruces NM 88003, USA}
 
 \IEEEauthorblockA{\IEEEauthorrefmark{2}Information Sciences (CCS-3), Los Alamos National Laboratory, Los Alamos NM 87544, USA}

  \IEEEauthorblockA{\IEEEauthorrefmark{3}Computational Earth Sciences (EES-16), Los Alamos National Laboratory, Los Alamos NM 87544, USA}}

\IEEEtitleabstractindextext{
\begin{abstract}
Quantum computing is an emerging topic in engineering that promises to enhance supercomputing using fundamental physics.  In the near term, the best candidate algorithms for achieving this advantage are variational quantum algorithms (VQAs). We design and numerically evaluate a novel ansatz for VQAs, focusing in particular on the variational quantum eigensolver (VQE). As our ansatz is inspired by classical multigrid hierarchy methods, we call it ``multigrid'' ansatz. The multigrid ansatz creates a parameterized quantum circuit for a quantum problem on $n$ qubits by successively building and optimizing circuits for smaller qubit counts $j < n$, reusing optimized parameter values as initial solutions to next level hierarchy at $j+1$. We show through numerical simulation that the multigrid ansatz outperforms the standard hardware-efficient ansatz in terms of solution quality for the Laplacian eigensolver as well as for a large class of combinatorial optimization problems with specific examples for MaxCut and Maximum $k$-Satisfiability. 
Our studies establish the multi-grid ansatz as a viable \textcolor{black}{method for improving the performance of variational quantum eigensolvers.}
\end{abstract}
}
\maketitle
\IEEEdisplaynontitleabstractindextext

{\section{Introduction}\label{sec:introduction}} 
Variational quantum algorithms are hybrid methods for gate-based quantum computers that combine quantum and classical techniques.   By partially offloading error-prone quantum processing to classical algorithms, these methods present the best opportunity for achieving quantum advantage in the Noisy Intermediate Scale Quantum (NISQ) era.   Due to this potential for impact, researchers continue to look for improvements to variational quantum algorithms. 

The VQE algorithm was proposed first by Peruzzo et al.~\cite{peruzzo2014variational}.  Following on this, a variational quantum algorithm for solving the Poisson equation was given by Sato et al.~\cite{sato2021variational}. This work was further developed on by Liu et al.~\cite{liu_vqa} who improved the number of measurements required from exponential to linear (making it practical) and connected it with QAOA.  A comprehensive review of the VQE space is given by Tilley et al.~\cite{tilly2022variational}, and a gentle introduction by Abhijith et al.~\cite{abhijith2018quantum}.

Other variants of VQE have been proposed, such as ADAPT-VQE~\cite{grimsley2019adaptive}.  This algorithm iteratively produces larger ans\"atze by Trotterization to solve chemical simulation problems.  Our work differs from this in that we aim for problems where there is a natural refinement structure, and especially \textbf{NP}-hard optimization problems such as MaxCut. \textcolor{black}{In addition, our approach generalizes a simple case of classical mesh refinement, where setting new parameters to zero corresponds to the output of the previous refinement layer.}  Another variant in this vein of ``growing ans\"atze" is Layer VQE~\cite{liu2022layer}, which increases the depth of its ansatz at each stage.  This method grows each successive ansatz in a completely ``different direction" than our work does, but is applied to combinatorial optimization problems.

In this paper, we introduce multigrid  methods for the variational quantum eigensolver (VQE) and test them against other quantum approaches and classical algorithms.  Multigrid methods use hierarchies of discretizations to solve problems (and especially differential equations) by moving from coarser to finer grids.  This process, known as mesh refinement, is the algorithm for which we propose a variational quantum counterpart.  Also, there is much potential for future work in this area: for example, more sophisticated multigrids and connections with other algorithms such as variational linear systems\cite{bravo2019variational}.

Multigrid methods have been applied to quantum annealers by Illa and Savage~\cite{illa2022basic} \textcolor{black}{and Lubasch et al. adapted finite-difference methods to a VQA for nonlinear PDEs~\cite{mg-vqa}}.  However, we are not aware of any published work developing multi-grid methods \textcolor{black}{(in particular, mesh refinement)} for gate-based quantum computing architectures.  We propose methods for platforms with all-to-all qubit connectivity, opening the pathway for future work to optimize these for other topologies.

\begin{figure*}
    \tikzset{every picture/.style={line width=0.75pt}}       
    \centering
    \begin{tikzpicture}[x=0.75pt,y=0.75pt,yscale=-1,xscale=1]

\draw   (252.15,85.69) .. controls (252.15,75.92) and (260.07,68) .. (269.84,68) -- (595.46,68) .. controls (605.23,68) and (613.15,75.92) .. (613.15,85.69) -- (613.15,138.76) .. controls (613.15,148.53) and (605.23,156.45) .. (595.46,156.45) -- (269.84,156.45) .. controls (260.07,156.45) and (252.15,148.53) .. (252.15,138.76) -- cycle ;
\draw   (249.05,207.65) .. controls (249.05,188.21) and (264.81,172.45) .. (284.25,172.45) -- (578.34,172.45) .. controls (597.78,172.45) and (613.54,188.21) .. (613.54,207.65) -- (613.54,313.25) .. controls (613.54,332.69) and (597.78,348.45) .. (578.34,348.45) -- (284.25,348.45) .. controls (264.81,348.45) and (249.05,332.69) .. (249.05,313.25) -- cycle ;
\draw   (261.15,274.54) .. controls (261.15,268.42) and (266.12,263.45) .. (272.24,263.45) -- (591.06,263.45) .. controls (597.18,263.45) and (602.15,268.42) .. (602.15,274.54) -- (602.15,307.81) .. controls (602.15,313.93) and (597.18,318.9) .. (591.06,318.9) -- (272.24,318.9) .. controls (266.12,318.9) and (261.15,313.93) .. (261.15,307.81) -- cycle ;
\draw  [fill={rgb, 255:red, 0; green, 0; blue, 0 }  ,fill opacity=1 ] (185.09,185.04) .. controls (185.59,229.24) and (209.32,264.81) .. (238.08,264.48) -- (238.26,280.11) .. controls (209.49,280.43) and (185.77,244.87) .. (185.27,200.66) ;\draw  [fill={rgb, 255:red, 0; green, 0; blue, 0 }  ,fill opacity=1 ] (185.27,200.66) .. controls (184.89,167.84) and (197.43,139.49) .. (215.68,126.93) -- (215.74,132.14) -- (236.35,112.22) -- (215.45,106.1) -- (215.51,111.3) .. controls (197.25,123.86) and (184.72,152.22) .. (185.09,185.04)(185.27,200.66) -- (185.09,185.04) ;
\draw   (5,83.89) .. controls (5,43.63) and (37.63,11) .. (77.89,11) -- (571.26,11) .. controls (611.52,11) and (644.15,43.63) .. (644.15,83.89) -- (644.15,302.56) .. controls (644.15,342.82) and (611.52,375.45) .. (571.26,375.45) -- (77.89,375.45) .. controls (37.63,375.45) and (5,342.82) .. (5,302.56) -- cycle ;
\draw  [fill={rgb, 255:red, 0; green, 0; blue, 0 }  ,fill opacity=1 ] (406,164) -- (412.79,164) -- (412.79,140) -- (426.36,140) -- (426.36,164) -- (433.15,164) -- (419.57,180) -- cycle ;
\draw  [fill={rgb, 255:red, 0; green, 0; blue, 0 }  ,fill opacity=1 ] (392,387) -- (398.79,387) -- (398.79,363) -- (412.36,363) -- (412.36,387) -- (419.15,387) -- (405.57,403) -- cycle ;
\draw  [fill={rgb, 255:red, 0; green, 0; blue, 0 }  ,fill opacity=1 ] (233.15,378) -- (226.36,378) -- (226.36,402) -- (212.79,402) -- (212.79,378) -- (206,378) -- (219.57,362) -- cycle ;

\draw (286.84,74) node [anchor=north west][inner sep=0.75pt]   [align=left] {\underline{Algorithm 3: Multigrid Refinement Step}\\\\1. Add a new qubit to define a refined problem\\2. Entangle that qubit with CZ and RY gates};
\draw (275.84,180.45) node [anchor=north west][inner sep=0.75pt]   [align=left] {\underline{Algorithm 2: Variational Quantum Eigensolver}\\\\1. Set new angles to 0 and append to old angles\\2.\textbf{ while} no convergence \textbf{and} $\textit{iter} < \textit{max\_iter}$,\textbf{	do}:\\
\hspace{1cm}Apply Algorithm 1};
\draw (277.15,269.5) node [anchor=north west][inner sep=0.75pt]   [align=left] {\underline{Algorithm 1: VQE Cost Function}\\1. Prepare trial wavefunction with current angles\\2. Estimate the eigenvalue with \textit{num\_shots} shots};
\draw (277.15,321.5) node [anchor=north west][inner sep=0.75pt]   [align=left] {\hspace{1cm}Update angles according to output};

\draw (65,176) node [anchor=north west][inner sep=0.75pt]   [align=left] {prior stage's\\angles, eigenvalue};

\draw (30,216) node [anchor=north west][inner sep=0.75pt]   [align=left] {\textit{(repeat until $max\_iter$ is}\\\textit{reached, or convergence)}};
\draw (49,37) node [anchor=north west][inner sep=0.75pt]   [align=left] {\underline{Algorithm 4: The Multigrid Variational Quantum Eigensolver}};
\draw (50,379) node [anchor=north west][inner sep=0.75pt]   [align=left] {inputs: ansatz, hamiltonian,\\num\_shots, max\_iter, ...};
\draw (425,380.03) node [anchor=north west][inner sep=0.75pt]   [align=left] {outputs: final angles, eigenvalue,\\optionally each stage's data};

\end{tikzpicture}
    \caption{A process diagram for the Multigrid VQE.  When one level of the Hamiltonian is optimized, refine the problem and use the previous level's angles to optimize.  This process proceeds until refinement produces the original Hamiltonian.}
    \label{fig:process}
\end{figure*}

In Section~\ref{sec:background}, we introduce variational quantum algorithms more thoroughly and explain two important examples.  We also discuss hardware-efficient quantum ans\"atze, which are used as a benchmark later on.  Section~\ref{sec:multigrid} discusses our methods, including Multigrid Hierarchies of Variational Ans\"atze and the Multigrid VQE.  In Section~\ref{sec:laplacian}, we discuss how to implement our methods for discrete Laplacian energy simulation, and our results for that problem.  Then, in Section~\ref{sec:combinatorial}, we look at \textbf{NP}-hard Maximum Cut problem on graphs.  
Section~\ref{sec:conclusion} concludes with a brief recap of the paper, and directions for future work.

\vspace{0.4cm}

\section{Background}

\textcolor{black}{For a general overview of quantum optimization, we recommend references~\cite{abbas2023optimization} and~\cite{symons2023optimization}.  The rest of this section will focus on variational quantum algorithms, and particularly our description of the variational quantum eigensolver.}

\label{sec:background}
\subsection{Variational Quantum Algorithms}
\textcolor{black}{Variational quantum algorithms (VQAs) are an active area of research in quantum information science, with the potential for achieving quantum advantage on near-term quantum computers~\cite{cerezo2021variational}. } They center on a parameterized quantum circuit $U(\theta_1,\theta_2,\ldots,\theta_n)$ called an \textit{ansatz} which prepares a trial wavefunction.  This quantum state is measured in various ways, and those measurements are used to classically calculate a score.  This score is fed to a classical optimization algorithm to update the parameters, and this optimization loop proceeds until convergence or a maximum number of iterations has been reached. At this point the angles are finalized and the corresponding state can be prepared at will and used for whatever purpose is desired, or the final score can be returned as output.

\begin{figure*}
\label{refinement_layer}
    \input refinement_layer
\end{figure*}

Important VQAs include the Variational Quantum Eigensolver (VQE) for finding the ground state of a Hamiltonian~\cite{peruzzo2014variational}, and the Quantum Approximate Optimization Algorithm (QAOA) for combinatorial optimization problems~\cite{farhi2014quantum}. In the long term, VQE has the potential to find  applications to chemistry, and QAOA to become a serious competitor for real-world combinatorial optimization applications.  As a result, considerable effort has been spent developing methods for these VQAs, and studying how different approaches affect performance across different classes of problems.  
\textcolor{black}{For example, ore sophisticated variants have been proposed, such as filtered VQE, warm-start QAOA, and recursive QAOA~\cite{f-vqe, ws-qaoa, r-qaoa}}.

These hybrid approaches allow for shallower and more robust circuits, offloading quantum complexity into more manageable classical overhead.  However, such methods come with substantial uncertainty; no speedup has been proven for VQAs, and their cost landscapes are susceptible to \textit{barren plateaus}~\cite{cerezo2021cost} where classical optimizers get caught descending into infinitely vanishing (or, in the case of maximization, infinitely growing) regions. However, preliminary experimental work by Boulebnane and Montenaro~\cite{boulebnane2022solving} and Golden et al.~\cite{golden2023numerical} suggest the potential for QAOA advantage.  Prior work by Farhi et al.~\cite{farhi2015quantum} on Max-$k$XOR had achieved a better guaranteed bound than the best known classical methods at the time, but better classical methods were subsequently found  by Barak~et~al.~\cite{barak2015beating}.

\subsection{Relevant VQA Methods} 
\subsubsection{Variational Quantum Eigensolver}

Given a Hamiltonian 
\begin{equation}\label{pauli_decomposition}H=\sum_i W_i^\dagger \alpha_i P_i W_i,\end{equation}
represented as a sum of unitary-transformed weighted Pauli strings (as in Equation~\eqref{pauli_decomposition}, where $W_i$ are unitary and $\alpha_i$ are weights), prepare a trial state $U(\theta_1,\theta_2,\ldots,\theta_n)$ using a VQA ansatz and calculate the weighted sum of measurements $M_i$ of each Pauli string $P_i$ on the corresponding state $W_iU(\theta_1,\theta_2,\ldots,\theta_n)$. Repeating this procedure many times and averaging yields an approximation of the Rayleigh quotient $\bra{U(\theta)}H\ket{U(\theta)}/\braket{U(\theta)|U(\theta)}$, which is an upper bound on the Hamiltonian ground state energy $\lambda_0(H)$.  (Note the denominator of this quotient is immaterial since quantum states have unit norm, and averaging works because the numerator is an expectation value.) Given an ansatz which can find this state and an effective optimization strategy, variationally minimizing this term can bring us arbitrarily close to (given enough shots) or even saturate this bound.  \textcolor{black}{Note that everything in this section also applies to Hamiltonians defined as weighted sums of Pauli strings without unitary transformation, by setting $W_i$ to the corresponding identity matrix for each term in Eq.~\ref{pauli_decomposition}.}

The VQE algorithm works like this, and can be seen as a quantum analogue of the classical Ritz method~\cite{ritz1909neue}. The absolute error scales down as the number of shots increases; heuristically, $\mathcal{O}(1/\varepsilon^2)$ shots are required~\cite{peruzzo2014variational} to obtain error within $\varepsilon$.  However, this result relies on various assumptions about the robustness of the ansatz and the well-suitedness of the classical optimization routine for the energy landscape.  Algorithm~\ref{vqecost} gives the VQE cost function computing the Rayleigh quotient for a given set of angles.  It does this by measuring each term in the decomposition of the  target Hamiltonian many times, and averaging the measurement results.   Algorithm~\ref{vqe} uses this function in the overall logic of the VQE algorithm, by varying  and optimizingthe ansatz parameters until either a maximum number of iterations is reached or convergence is achieved.  Figure~\ref{fig:process} shows our modified VQE, involving also Algorithms~\ref{alg:multigrid_vqe} and~\ref{alg:multigrid_refinement}  (defined in Section~\ref{sec:multigrid}).  In order to measure arbitrary Pauli strings, this involves applying change of basis gates: apply $H$ to a qubit to measure in the $X$ basis, and apply $S^\dagger H$ to measure in the $Y$ basis. Of course, decomposing $H$ according to Equation~\eqref{pauli_decomposition} can be nontrivial, as can the efficient implementation of arbitrary unitaries $W_i$. In particular, representing an $n$-qubit Hamiltonian $H$ as a sum of Pauli strings takes $4^n$ terms in general.  Thus it is important to measure in more convenient bases.   For example, Liu et al.~\cite{liu_vqa} use the operator basis $\mathcal{B}=\{I, \sigma_+, \sigma_-\}$, where $\sigma_+=\ket{0}\bra{1}$ and $\sigma_-=\ket{1}\bra{0}$, in their VQA for solving the Poisson equation.
\begin{algorithm}[b!]
    \caption{The VQE Cost Function (vqe\_cost)}\label{vqecost}
\ \newline\textbf{input: } Ansatz $U$, Hamiltonian $H$, Angles $\theta_1,\theta_2,\ldots,\theta_m$, Number of shots $num\_shots$.

\textbf{output: } Estimated eigenvalue $\lambda$.\newline

\textit{estimated\_eigenvalue} = $0$;

$W_i$, $P_i$, $\alpha_i$ = Decompose Hamiltonian according to Eq.~\ref{pauli_decomposition};

\textbf{for} $0\leq i<n$, \textbf{do}

\hspace{.5cm} \textit{term\_eigenvalue} = $0$;

\hspace{.5cm} // $H.num\_qubits$ is the number of qubits in the system $H$

\hspace{.5cm} \textit{circ} = \textbf{new} QuantumCircuit($H.num\_qubits$)

\hspace{.5cm} // Appending Q1 to Q2 means performing Q2 then Q1

\hspace{.5cm} Append $U(\theta_1, \theta_2, \ldots, \theta_m)$ to 
\textit{circ};

\hspace{.5cm} Append $W_i$ to \textit{circ};

\hspace{.5cm} Add measurement gates for $P_i$ to \textit{circ};

\hspace{.5cm} \textbf{do} $num\_shots$ \textbf{times} 

\hspace{1cm} \textit{measure} = Execute \textit{circ} on quantum computer;

\hspace{1cm} \textit{test\_eigenvalue} = pow$(-1,$ hamming$($\textit{measure}$))$;

\hspace{1cm} \textit{term\_eigenvalue} += $\alpha_i *\textit{test\_eigenvalue}/ num\_shots$;

\hspace{.5cm} \textbf{end do}

\hspace{.5cm} \textit{estimated\_eigenvalue} += \textit{term\_eigenvalue};

\textbf{end for}

\textbf{return} \textit{estimated\_eigenvalue};
\end{algorithm}

\begin{algorithm}[b!]

\caption{The VQE Classical Loop (vqe)}\label{vqe}

\ \newline\textbf{input: }  Ansatz $U$, Hamiltonian $H$, Angles $\theta_1,\theta_2,\ldots, \theta_m$, Numbers $max\_iter$, $num\_shots$.

\textbf{output: } Estimated eigenvalue $\lambda$, Final angles $\bm{\theta}_i'$, Optional optimizer information (e.g. number of function evaluations).\newline

$num\_iter = 0$;

$\textit{estimated\_eigenvalue} = \text{vqe\_cost}(U, H, \bm{\theta}_i, {num\_shots})$

\textbf{while} no convergence \textbf{and} $num\_iter<max\_iter$, \textbf{do}

\hspace{.5cm} $\theta_1,\theta_2,\ldots,\theta_m =\textnormal{optimize\_angles}(estimated\_eigenvalue,$ 

\hspace{1cm} $\theta_1,\theta_2,\ldots,\theta_m$);  // Using a classical optimizer.

\hspace{.5cm} $\textit{estimated\_eigenvalue} = \text{vqe\_cost}(U, H, \theta_i, {num\_shots})$;

\hspace{.5cm} $num\_iter$ += $1$;

\textbf{end while}

\textbf{return} \textit{estimated\_eigenvalue}, $\bm{\theta}_i$, optional info (such as number of optimizer calls or other classical optimizer results);

\end{algorithm}

\subsubsection{Quantum Approximate Optimization Algorithm}

\begin{figure*}[t]
    \input efficientsu2_circuit
\end{figure*}

Given a cost function $C$ on bitstrings, we can encode the corresponding optimization problem in a diagonal matrix
\begin{equation}\label{diagonal_cost}H_C=\sum_{i\in 2^n} C(i)\mathbf{e}_i,\end{equation}
 and approximately optimize $C$ using a discretization and Trotterization of the Quantum Adiabatic Algorithm~\cite{farhi2000quantum, farhi2014quantum}.  This method is called QAOA and was first applied to the NP-complete Maximum Cut problem (MaxCut) for the maximum number of edges between two complementary vertex subsets of a given graph $G$.  It is worth noting that this diagonal Hamiltonian $H_C$ can be fed into VQE if a suitable representation is known (see Section~\ref{sec:combinatorial} for an example).
 
\subsubsection{Hardware-Efficient Ans\"atze}\label{sec:hardware_efficient_ansaetze}

In order to avoid excessive noise in a VQA, efficient operations and low circuit depth are desired.  Hardware-efficient ans\"atze use native gates (such as nearest-neighbor interactions for superconducting qubits) to form a robust constant-depth circuit for preparing trial states~\cite{kandala2017hardware}.  These have been shown to suffer from barren plateaus at long depths, but to have the potential for quantum advantage when short~\cite{leone2022practical}.  Hardware efficient ans\"atze provide reasonable benchmarks for testing VQAs across different classes of problems due to their general nature and resultingly broad applicability.

\begin{figure*}
\centering
  \begin{subfigure}{0.25\textwidth}
    \includegraphics[page=1,width=1\textwidth]{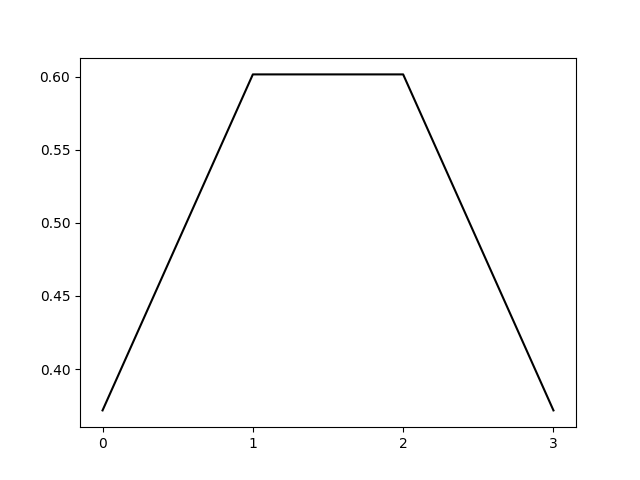}
    \caption{two qubits}
  \end{subfigure}\hfill%
  \begin{subfigure}{0.25\textwidth}
    \includegraphics[page=1,width=1\textwidth]{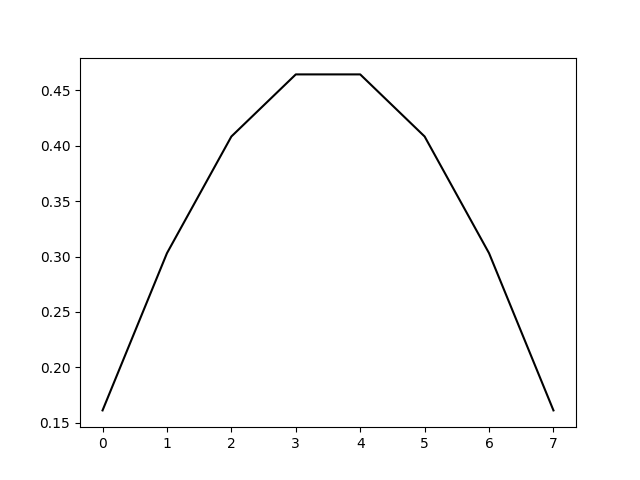}
    \caption{three qubits}
  \end{subfigure}\hfill%
  \begin{subfigure}{0.25\textwidth}
    \includegraphics[page=1,width=1\textwidth]{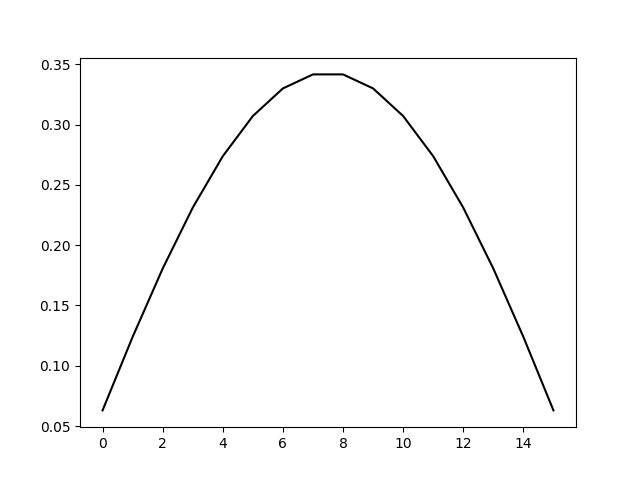}
    \caption{four qubits}
  \end{subfigure}\hfill%
  \begin{subfigure}{0.25\textwidth}
    \includegraphics[page=1,width=1\textwidth]{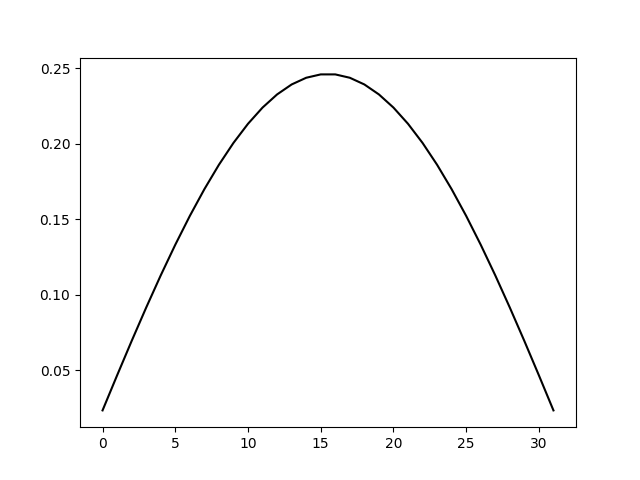}
    \caption{five qubits}
  \end{subfigure}%
  \caption{The smallest eigenvectors of $\nabla^2_D$ (for dimensions $m=2^2, 2^3, 2^4, 2^5$, respectively) form a refinement structure well-suited to multigrid methods. The $X$ axes represent computational basis states expressed as decimal integers, and the $Y$ axes show the amplitudes of the eigenvector in the respective entries. Ignoring a complex global phase, we assume entries in $\mathbb{R}_+$.
  }
  \label{fig:dirichlet_eigenvectors}
\end{figure*}

\begin{algorithm}[t!]

\caption{The Multigrid VQE (multigrid\_vqe)}\label{alg:multigrid_vqe}

\ \newline\textbf{input: } Seed ansatz $U$, Family of Hamiltonians $\{H_i\}$, Initial angles $\bm{\theta}_i$, Numbers $num\_shots$, $max\_iter$.

\textbf{output: } Estimated eigenvalue $\lambda$, Final angles $\bm{\theta}_i'$, Optional optimizer information and coarse stage data.\newline

$\lambda, \theta_i = \text{vqe}(U, H, \theta_1, \theta_2, \ldots, \theta_m, max\_iter, num\_shots)$;

\textbf{for} $U.num\_qubits< j <=H.num\_qubits$, \textbf{do}

\hspace{.5cm} $\theta_{m+1},\theta_{m+2},\ldots, \theta_{m+U.num\_qubits}=0,0,\ldots,0$;

\hspace{.5cm} $U_j = \text{multigrid\_refine}(U_{j-1})$; // With $U_{U.num\_qubits}$ = $U$
\hspace{.5cm} $\lambda, \theta_i = \text{vqe}(U_j, H_{j}, \theta_1, \theta_2, \ldots, \theta_{m}, \ldots, \theta_{m+U.num\_qubits},$

\hspace{1cm} $max\_iter, num\_shots)$;

\hspace{.5cm} $m=m+U.num\_qubits$;  // For the next iteration.

\textbf{end for}

\textbf{return} $\lambda, \bm{\theta}_i$;

\end{algorithm}

\section{Multigrid VQE}
\label{sec:multigrid}

We employ an approach based on classical multigrid methods, where hierarchies of increasingly finer discretizations are used to solve problems, such as differential equations.  First solving a coarser problem with an initial variational ansatz $U(\theta_1,\theta_2,\ldots\theta_n)$, we save the optimized angles $\theta_i$ and use these to seed the next round.  In this subsequent round, we apply the ansatz $U$ (which we call the \textit{seed ansatz}) then entangle it to a new qubit as shown in Figure~\ref{fig:multigrid_refinement}.  This method preserves the multigrid intuition because setting the initial angles for the \texttt{RY} gates entangling the new qubit to zero puts the system in the state $U(\theta_1,\theta_2,\ldots,\theta_n)\ket{00\ldots0}\otimes \ket{+}$, which corresponds to a constant interpolation of the state from the previous round.  That is, if $U(\theta_1,\theta_2,\ldots,\theta_n)\ket{00}=\ket{\psi}$, then 
\begin{equation}
    U(\theta_1,\theta_2,\ldots,\theta_n)\ket{00}\otimes\ket{+}=\frac{\ket{\psi}\ket{0}+\ket{\psi}\ket{1}}{\sqrt{2}},
\end{equation}
the equal superposition of adding a $\ket{0}$ qubit and adding a $\ket{1}$ qubit.  This occurs since neutralizing the $Y$ rotations on the added qubit causes all $\texttt{CZ}$ gates to annihilate, yielding simply a Hadamard operation returning the plus state on that qubit.  We continue adding qubits until we reach the target system.

The two-qubit gate cost of this method is only quadratic in the number of qubits, keeping it reasonable for use on NISQ devices.  In state preparation problems, this makes the gate cost polylogarithmic in the number of amplitudes (assuming polynomial circuits for any unitary transforms in the Pauli decomposition of the problem Hamiltonian, Equation~\eqref{pauli_decomposition}).  This will be relevant in the next section, where we use Multigrid VQE to prepare the ground state of a discrete Laplacian Hamiltonian.   The overall idea is that for problems with a hierarchical structure, information from easier, lower stages can be used to improve the performance of harder, higher stages.  There is no particular reason this process has to proceed one qubit at a time, either; future work could explore the integration of multiple new qubits at once, or designs that use fewer entangling gates in each consecutive layer for implementations on qubit topologies, like with IBM's heavy-hex lattice architecture, that lack all-to-all connectivity. 

\begin{algorithm}[t!]

\caption{Multigrid Refinement (multigrid\_refine)}\label{alg:multigrid_refinement}

\ \newline\textbf{input: } A parameterized quantum circuit $U_n$ with $n$ qubits.

\textbf{output: } A parameterized quantum circuit $U_{n+1}$ with $n+1$ qubits.\newline

\textit{circ} = new QuantumCircuit($U.num\_qubits+1$);

Append $U$ to the first $U.num\_qubits$ qubits of \textit{circ};\newline 

// We assume that qubit addresses start at zero.

Append a Hadamard gate to qubit $n$ of \textit{circ};\newline

\textbf{for} $i<n$, \textbf{do}

\hspace{.5cm} Append a $CZ$ gate between qubits $i$ and $n$ of \textit{circ};

\hspace{.5cm} Add a new variational parameter $p_{m+i}$ to \textit{circ};

\hspace{.5cm} Append the gate $RY(p_{m+i})$ to qubit $n$ of \textit{circ};

\hspace{.5cm} Append a $CZ$ gate between qubits $i$ and $n$ of \textit{circ};

\textbf{end do}

\textbf{return} \textit{circ};

\end{algorithm}

\begin{figure}
\input shift_circuit
\end{figure}

Algorithm~\ref{alg:multigrid_vqe} uses multigrid refinement to perform what we call multigrid VQE, assuming a hierarchy of properly represented Hamiltonians.  The idea is to solve progressively larger Hamiltonians using information about each previous stage's angles until you reach your target Hamiltonian. Based on the design of the multigrid refinement function, we can think of this as an analogue of classical mesh refinement with constant interpolation.  One drawback of this approach is that the maximum number of total optimizer iterations goes to infinity as the size of the problem scales up.  This can be addressed by choosing larger seed ans\"atze for larger problems.  In this work, we do not simulate large enough $n$ for this to become an issue, so we use the Efficient SU(2) seed ansatz to standardize our experiments. \textcolor{black}{Given an ansatz with a total of $n_{s.a.}$ parameters, the Multigrid Hierarchy has }
\begin{equation}\label{quadratic_gatecost}n_{s.a.}+\sum_{i=m}^{n-1} i= n_{s.a.}+\frac{n^2-n-m^2+m}{2}\end{equation}

\noindent parameters.  Now $m=2$ and $n_{s.a.}=16$ for the Efficient SU(2) ansatz with 2 qubits, so at the $n=10$ layer this approach uses $60$ parameters. \textcolor{black}{The number of parameters therefore increases only quadratically in the number of qubits, and polylogarithmically in the refinement depth provided by the number of statevector amplitudes.  Further, assume unitaries in the Pauli decomposition  (Equation~\eqref{pauli_decomposition})  of the problem Hamiltonian are implementable in polynomial depth.  Then each VQE circuit has depth $O(n^2)+O(\textnormal{poly}(n))=O(\textnormal{poly}(n))$ in the number of qubits, and is $O(\textnormal{polylog}(n))$ in the refinement depth.}

\begin{figure*}[t]
\centering
\begin{subfigure}{0.49\textwidth}
    \includegraphics[page=1,width=1\textwidth]{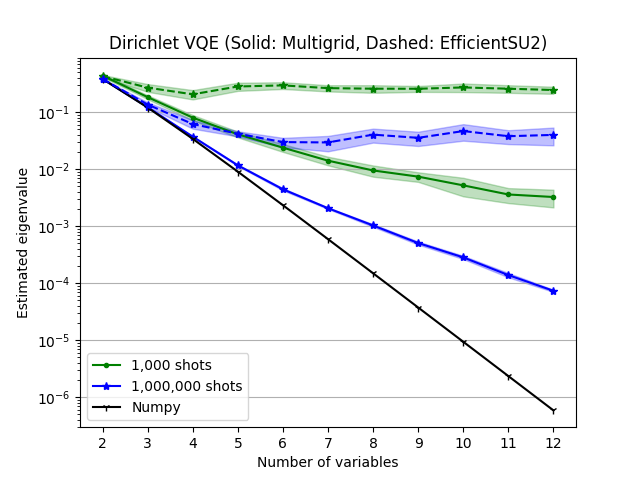}
    \caption{} \label{fig:Dirichlet_a}
  \end{subfigure}\hfill%
  \begin{subfigure}{0.49\textwidth}
     \includegraphics[page=1,width=1\textwidth]{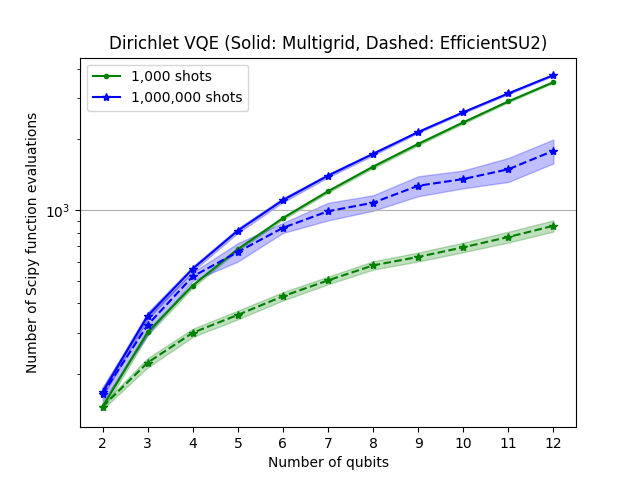}
    \caption{} \label{fig:Dirichlet_b}
  \end{subfigure}%
\caption{The Multigrid outperforms the static Efficient SU(2) ansatz for Dirichlet Laplacian VQE. In \ref{fig:Dirichlet_a}, we plot the estimated eigenvalues \textcolor{black}{(not the accuracy in eigenvalues)} and observe that the Multigrid continues improving while the Efficient SU(2) ansatz flatlines quickly. As expected, increasing the number of shots improves performance. In \ref{fig:Dirichlet_b}, we observe that Multigrid VQE uses more optimizer calls rather than giving up quickly, and has less variance in the number of optimizer calls.} \label{Dirichlet}
\end{figure*}

To seed our multigrids, and as a benchmark for our methods, we use a family of hardware-efficient ans\"atze called Efficient SU(2) ans\"atze.  These are implemented in Qiskit as the class \texttt{qiskit.circuit.library.EfficientSU2}, and are primarily defined by a number of repetitions of single qubit  gates---that is, $SU(2)$ unitaries---followed by $CX$ entanglements as shown in Figure~\ref{fig:efficient_ansatz}.  This family has many of the hardware-efficient benefits discussed in Section~\ref{sec:hardware_efficient_ansaetze} such as only nearest-neighbor interactions and low depth suggesting the potential for quantum advantage.  We construct Efficient SU(2) circuits using the default arguments from Qiskit including the reverse linear entanglement pattern and the $\{RY,RZ\}$ gate set, but future work could explore how different choices for seed ansatz arguments affect multigrid performance on particular problems.  \textcolor{black}{For example, more highly expressible seed ansatzes should have better performance; and other factors such as size or entanglement patterns could also affect performance.} Our graphs are generated with $95$th-percentile confidence intervals calculated by $z$-score. This is not meant to imply that our data is normally distributed, \textcolor{black}{but to represent the uncertainty in the \textit{mean} of the samples.} Simulations were done in Python 3, where quantum circuits were implemented in Qiskit~\cite{Qiskit} and we used Scipy's optimization libraries.  Given data from these simulations, Numpy and PyPlot were used for analysis and plotting.  In the next two sections, we compare performance between Efficient SU(2)-seeded Multigrid VQE, Efficient SU(2) VQE, and some classical methods: Numpy for eigenvalue problems and greedy, brute-force methods for combinatorial optimization.

\section{Laplacian Eigensolver}
\label{sec:laplacian}

We use multigrid VQE to approximately prepare the ground state of (and estimate the corresponding ground energy for) the discrete one-dimensional Laplacian with constant zero Dirichlet boundary conditions.  By the second difference methods expanded on in Appendix~\ref{appendix}, this Hamiltonian is the $m\times m$ Toeplitz tridiagonal matrix given by

\[\nabla^2_D=\begin{pmatrix}
2 & -1 &  &  &  &  &  \\
-1 & 2 & -1 &  &  &  &  \\
 & -1 & 2 & -1 &  &  &  \\
 &  & \ddots & \ddots & \ddots &  &  \\
 &  &  & -1 & 2 & -1 & \\
 &  &  &  & -1 & 2 & -1 \\
 &  &  &  &  & -1 & 2 
\end{pmatrix}\]

\noindent which Sato et al.~\cite{sato2021variational} give for the case where $m$ is any power of two in a more convenient form as
\begin{equation}
    \nabla^2_D=S+P^\dagger SP+P^\dagger(I_0^{\otimes n-1}\otimes X)P.
\end{equation}
 
 Here $P$ is the cyclic shift unitary described in more detail below, $S=I^{\otimes n-1}\otimes(I-X)$ and $I_0=\ket{0}\bra{0}=(I+Z)/2$.  Expanding these definitions allows us to express the Dirichlet Hamiltonian purely in terms of unitary-transformed Pauli strings, allowing us to implement VQE for this problem.  Specifically, we have unshifted (or periodic) terms $S=1-X_0$ and cyclically shifted (by $P$) terms $S=1-X_0$ and
 \[2^{-n+1} (1+Z_0+Z_1Z_0+Z_2Z_0+\cdots + Z_{n-2}Z_{n-1}\cdots Z_1Z_0)X_{n-1},\]
 which can be optimized~\cite{liu_vqa} for efficiency in any practical implementation. However, for our purposes we only need the theoretical performance of the algorithm.  To speed up simulations, we use the fact that these Pauli strings all commute to calculate the whole expectation value in just one measurement.
 
 For an increment circuit implementing the cyclic shift operation $P$, see Figure~\ref{shift_circuit}.  \textcolor{black}{As referenced in the caption there, this circuit can be implemented with a linear overhead in number of gates.} The important part for us is that it acts on computational basis vectors by sending big-endian computational basis states $\ket{b}$ to $\ket{\texttt{rem}(b+1,2^n)}$.  For the simpler case with periodic boundary conditions, we follow Sato et al. in measuring the simpler Hamiltonian $\nabla^2_P=S+P^\dagger SP$. (For an explicit matrix representation, see Appendix~\ref{appendix}.)  This is convenient for representing the Dirichlet Hamiltonian as 
$\nabla^2_D=\nabla^2_P+P^\dagger (I_0^{\otimes n-1}\otimes X)P$, but preparing the eigenstate of $\nabla_P$ (and thus calculating its ground state energy) is not an interesting application of VQE since the state you try to get is simply the superposition of computational basis states.  This state can be prepared directly by a circuit applying a Hadamard gate to each qubit in the quantum register.

Ground state simulation for the discrete Dirichlet Laplacian, on the other hand, is more interesting.  The states are less trivial than for the discrete Periodic Laplacian, but they still provide some structure to work with.  In fact, the problem is well-suited to multigrid methods. This is because the ground states of these Hamiltonians seem to converge smoothly.  Seeing this requires viewing them as functions from the computational basis states to $\mathbb{R}_+$. The corresponding graphs visibly form a series of increasingly smooth discrete functions with a more or less parabolic shape. This interesting structure motivates our Multigrid VQE methods and provides a good test case.  For an illustration of this refinement structure and further commentary, see Figure~\ref{fig:dirichlet_eigenvectors}.  Future work could try to produce a circuit implementing a more linear interpolation better suited to this problem than our constant interpolation scheme.

\textcolor{black}{The circuits in this method are polylogarithmic in the size of the obtained eigenvector.  This follows from Equation~\eqref{quadratic_gatecost} together with how increment circuits implementing the shift operation can be constructed with polynomially many CNOTs.} This sort of benefit is common for VQE, and partially explains its popularity.  We implement Multigrid VQE for Laplacian ground state preparation in Qiskit~\cite{Qiskit}, and compare its estimated performance both with static application of the Efficient SU(2) ansatz and with a brute force method.  We seed the Multigrid Hierarchy with an Efficient SU(2) ansatz as a proof of concept.  This choice is somewhat arbitrary and does not come with a multigrid-related theoretical justification.  Future work could test other seed ans\"atze, and seed sizes.  In this case, we use an ansatz with two qubits since the discrete Dirichlet Laplacian is not defined on fewer than two qubits.  There is a one qubit analog~\cite{liu_vqa} given by the formula $A=2I-X$, but this has an uninteresting ground state, the plus state. Therefore $A$ is more analogous to $\nabla^2_P$ than to our target of $\nabla^2_D$.  However, the ground state energy of $\nabla^2_D$ vanishes quickly in the number of qubits and thus needs increasingly more shots to maintain its accuracy.  However, this only reflects the exponential increase of the size of the multigrid resolution, and the number of shots grows more reasonably in terms of that.

\subsection{Simulation results}

To create the plots in Figure~\ref{Dirichlet}, we generated the multigrid hierarchy of Dirichlet Laplacians from the two to the twelve qubit case.  Running each Hamiltonian through Multigrid VQE and Efficient SU(2) VQE, we  estimated eigenvalue and number of optimizer calls. The solid green line represents the $10^3$-shot Multigrid VQE, and the dashed green line is for the $10^3$-shot Efficient SU(2) VQE.  The solid blue line represents the $10^6$-shot Multigrid VQE, and the dashed blue line is for the $10^6$-shot Efficient SU(2) VQE.  We estimate the performance of our method on Dirichlet Hamiltonians with as many as~12 qubits, corresponding to preparing eigenstates with~4,096 amplitudes.  \textcolor{black}{These experiments are performed using Qiskit's Aer simulator, assuming no hardware noise.}

\begin{figure}
\centering
    \includegraphics[scale=0.5]{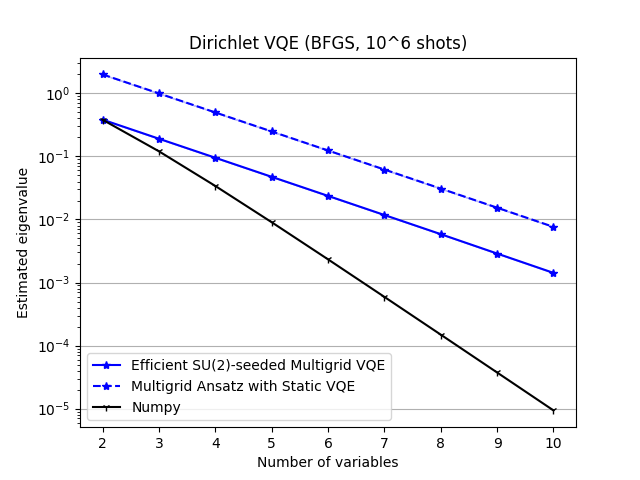}
  \caption{In the dashed line, we test BFGS with all zero angles at every stage.  In the solid line, COBYLA is used to seed a BFGS Multigrid VQE.  The apparent scaling advantage suggests that gradient methods can be used to preserve good solutions once COBYLA becomes unwieldy due to number of parameters. }
  \label{fig:BFGS}
\end{figure}

We observe significant improvement over the static Efficient SU(2) ansatz when using the dynamic Multigrid VQE.  This result suggests that the refinement structure of the problem is being taken advantage of. In Figure~\ref{fig:Dirichlet_a}, the static ans\"atze quickly flatline in performance, while the Multigrid VQE is able to squeeze out more accuracy.  Furthermore, increasing the number of shots for the Multigrid VQE roughly lines up with the heuristic prescribing $\mathcal{O}(1/\varepsilon^2)$ shots for error $\varepsilon$.  With $10^3$ shots it achieves error $10^{-2}$, and with $10^6$ shots, it eventually achieves error $10^{-3}$.  On the other hand, the static VQE does not even reach $10^{-1}$ accuracy with $10^3$ shots, or $10^{-2}$ accuracy with $10^6$ shots.  Interestingly, $10^6$ shot Multigrid VQE appears strikingly precise, producing very similar values in each test.  Also, there is a crossover point after which $10^3$ shot Multigrid VQE appears to outperform $10^6$ shot Efficient SU(2) VQE despite its having substantially more \textcolor{black}{parameters}.  In Figure~\ref{fig:Dirichlet_b}, we see that this method uses more function evaluations (optimizer calls) as a trade-off.  On the other hand, it shows how Multigrid VQE is able to take more time thoroughly exploring the parameter space rather than giving up too quickly.

\begin{figure*}
\centering
  \begin{subfigure}{0.49\textwidth}
    \includegraphics[page=1,width=1\textwidth]{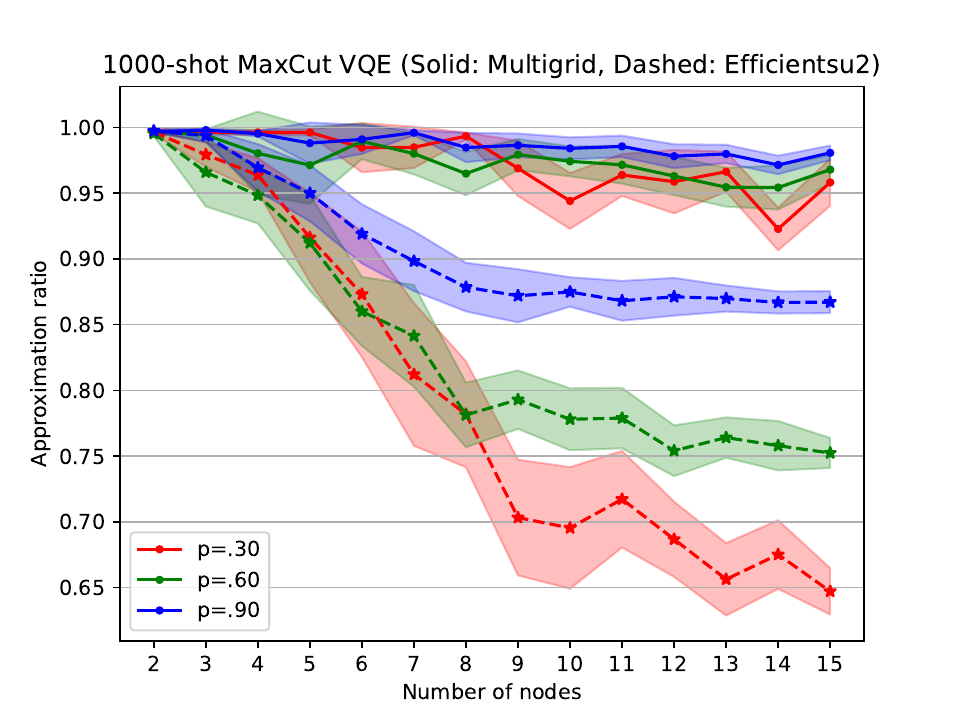}
    \caption{} \label{fig:MaxCut_a}
  \end{subfigure}\hfill%
  \begin{subfigure}{0.49\textwidth}
    \includegraphics[page=1,width=1\textwidth]{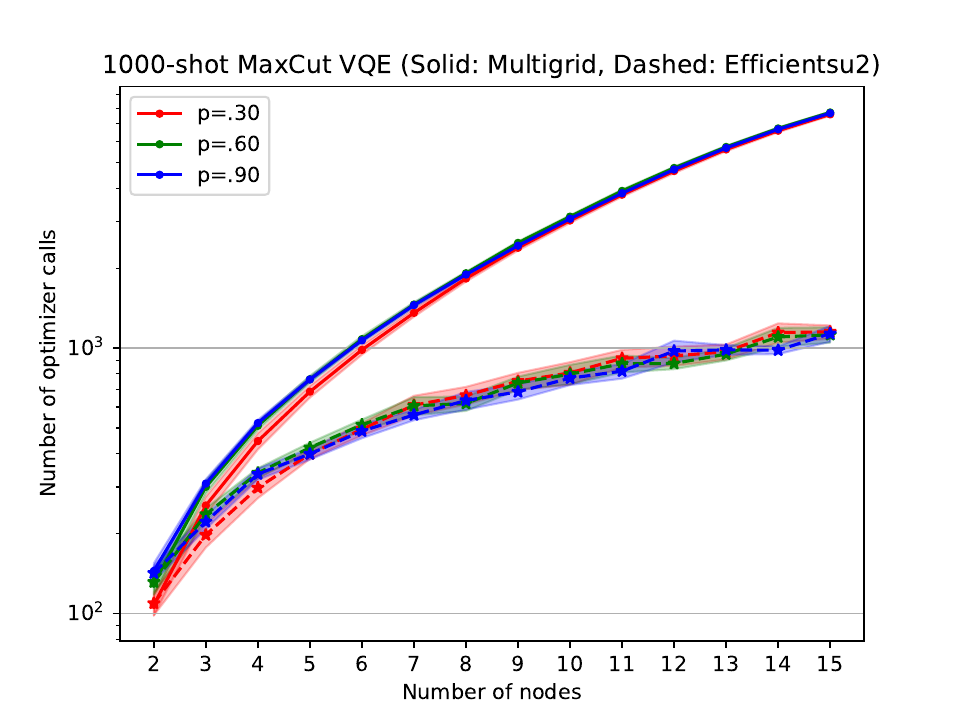}
    \caption{} \label{fig:MaxCut_b}
  \end{subfigure}%
  \caption{(a) The Multigrid VQE achieves better and more stable approximation values for MaxCut than the static Efficient SU(2) ansatz does.  Also, $p$ seems to affect Efficient SU(2) performance much more drastically.  (b) Multigrid VQE is able to use more optimizer calls than the Efficient SU(2) ansatz VQE, and experiences less variance in number of optimizer calls.}
  \label{fig:MaxCut}
\end{figure*}

Figure~\ref{fig:BFGS} demonstrates clearly how the Multigrid VQE method of saving angles from previous rounds can lead to better solutions.  The dashed line represents static application of the Multigrid Ansatz (i.e., using the ans\"atze produced by Algorithm~\ref{alg:multigrid_refinement} but with all angles initialized to zero rather than just the new ones).  In contrast, the solid line represents performing the Efficient SU(2)-seeded Multigrid VQE as defined in Algorithm~\ref{alg:multigrid_vqe}.  To draw inferences about this method's ability to keep producing better results when methods such as COBYLA (Constrained Optimization by Linear Approximation) become unwieldy for large systems beyond the reach of our simulations, we use BFGS and observe that the Multigrid VQE values are consistently better than their static counterparts.

\section{Combinatorial Optimization}
\label{sec:combinatorial}

An instance of a combinatorial optimization problem is a set $S\subseteq \{0,1\}^n$ of bitstrings together with a function $f:S\to\mathbb{R}$ called the cost function. When $S=\{0,1\}^n$ we call the problem unconstrained, and constrained otherwise.  A solution to such an instance is a bitstring $z\in S$ with $f(z)=\min_{z\in S}f(z)$. Combinatorial optimization problems are important in computer science, where they provide a framework for studying many classes of \textbf{NP} problems. In order to apply our multi-grid hierarchy ansatz to combinatorial optimization, we successively grow the problem instance by defining solution bitstring sets $S_j \subseteq \mathbf{2}^j$ and corresponding cost functions $f_j:S_j\to\mathbb{R}$ for $0<j\leq n$ with $S = S_n$ and $f = f_n$. The exact relationship of $f_i$ and $f_{i+1}$ depends on the nature of the problem; for graph problems, for example, we insert the $i+1$-st vertex and its corresponding edges that connect to the already inserted first $i$ vertices. We give two concrete examples below.  

\begin{figure*}
\centering
  \begin{subfigure}{0.49\textwidth}
    \includegraphics[scale=0.5]{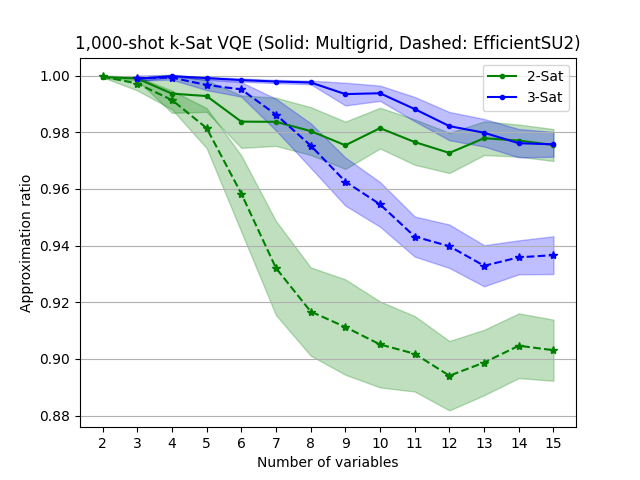}
    \caption{} \label{fig:ksat_a}
  \end{subfigure}\hfill%
  \begin{subfigure}{0.49\textwidth}
    \includegraphics[scale=0.5]{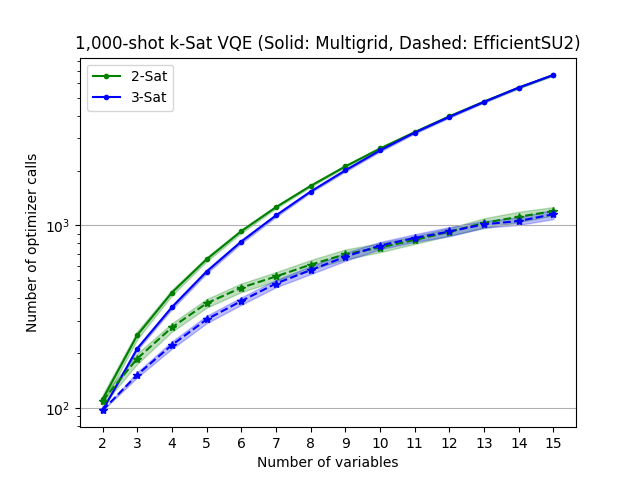}
    \caption{} \label{fig:ksat_b}
  \end{subfigure}%
  \caption{(a) The Multigrid VQE achieves better approximation values for hard instances of Max $2$ and $3$-Sat than the Efficient SU(2) ansatz. Furthermore, this performance seems to be consistent across the different values of  $k$. (b) The Multigrid VQE uses more total optimizer calls than the Efficient SU(2) ansatz, showing its better ability to explore the feasible state space.}
  \label{fig:ksat}
\end{figure*}

When a combinatorial optimization problem is defined using a diagonal encoding $H_C$ of the cost function as in QAOA, this Hamiltonian representation can be used to solve the problem with VQE because the target is either the minimum eigenvalue $\lambda_0(H_C)$ or the maximum,  $-\lambda_0(-H_C)$.  For unconstrained problems, this is all we need; and for constrained problems we can use penalty terms to enforce the constraints.

To demonstrate these methods and test the performance of Multigrid VQE in such a setting, we implement VQE for two unconstrained optimization problems: Maximum Cut and Maximum Exact $k$-Satisfiability.  These problems are both \textbf{NP}-hard, and future work could look into constrained versions of them such as Maximum Bisection.

\subsubsection{Maximum Cut}
Given a graph $G=(V,E)$, a cut on $G$ is a partition of $V$ into complementary subsets $U_1\sqcup U_2=V$.  The size of a cut is the number of edges between $U_1$ and $U_2$.  The Maximum Cut (MaxCut) problem is to find the maximum size among all cuts on $G$. We implement Multigrid VQE for MaxCut by minimizing the inverse MaxCut Hamiltonian representation given by the weighted sum of Pauli strings
\begin{equation}H_C=\frac{1}{2}\sum_{(v,w)\in E}\big(Z_vZ_w-1\big),\end{equation}
\noindent which requires only one measurement since all measurements are in the $Z$ basis.  These measurements commute so that it is enough to measure all qubits simultaneously and classically compute the $ZZ$ terms. Letting $n=|V|$, the complexity of computing the cost function from a measurement is then $\mathcal{O}(n^2)$, since $|E|\leq |E(K_V)|=(n^2-n)/2$. In fact, you need only measure qubits corresponding to nonzero degree nodes; but this does not affect scaling in general.

In order to use the Multigrid Hierarchy in this setting, we need a corresponding Subgraph Hierarchy to perform MaxCut on at each stage as we build the graph.  We choose to naively add nodes in the order they are labeled in the NetworkX library by which they are generated.  Generating Erdős-Rényi random graphs, we compare the  performance of Multigrid VQE to static application of the Efficient SU(2) ansatz.  To enable this, we compute the approximation ratio $C(z_\textnormal{obs})/C(z_\textnormal{opt})$, using a brute force method to compute $C(z_\textnormal{opt})$.  Here $z_\textnormal{obs}$ is the observed bitstring that we are computing the approximation ratio for and $z_\textnormal{opt}$ is an optimal bitstring.

\subsubsection{Maximum $k$-Satisfiability}

Given a propositional formula on $n$ variables, in conjunctive normal form with exactly $k$ variables in each of $m$ clauses, the Maximum Exact $k$-Satisfiability problem (MaxE$k$-Sat) asks how many clauses can be satisfied simultaneously.  We encode such formulae into Hamiltonians using a clause-by-clause decomposition.  Given a clause, we produce the $k$-local Hamiltonian $I^{\otimes k}-H$ where $H$ applies $\ket{0}\bra{0}$ to the qubits corresponding to true propositions, and $\ket{1}\bra{1}$ to the qubits corresponding to false ones.  This process, derived from the De Morgan duality, penalizes bitstrings satisfying the negation of the clause.

For example, the clause $x_2\vee \neg x_3$ maps to the $2$-local Hamiltonian $I^{\otimes 3}-I\otimes \ket{0}\bra{0}\otimes \ket{1}\bra{1}$.  The whole problem Hamiltonian is given by the sum of these Hamiltonians, for each clause in the formula.  To convert this Hamiltonian to the Pauli-$Z$ basis, we use the mappings $\ket{0}\bra{0}\to (I+Z)/2$ and $\ket{1}\bra{1}\to (I-Z)/2$.  Expanding out the resulting tensor product yields only $\mathcal{O}(m2^k)$ terms, so it is efficient in terms of $m$ and $n$.  This means that our example becomes 
\[I^{\otimes 3}-\frac{I^{\otimes 3}-I^{\otimes 2}\otimes Z+I\otimes Z\otimes I-I\otimes Z^{\otimes 2}}{4}.\]
All Pauli-$Z$ measurements commute, so we can optimize the circuit to only use one circuit execution per shot.  Furthermore, we only need to measure qubits corresponding to variables that occur in some term (although this does not matter practically).

We produce a Multigrid Hierarchy for this problem by adding variables in the order they are generated by our code.  When a variable is added, all clauses including that variable or variables already added are admitted.  However, we do not admit any clause involving any variable not yet added.  

We compare Multigrid Random MaxE$k$-Sat VQE to static Efficient SU(2) VQE, using a brute force algorithm to compute exact solutions for determining approximation ratios.  Following Golden et al.~\cite{golden2023quantum}, we generate hard instances beyond the difficulty phase transition by setting $m=3n$ for the $k=2$ case and $m=6n$ for the $k=3$ case.

\subsection{Simulation Results}
\textcolor{black}{As in the previous section, we run our tests with Qiskit's Aer simulator, assuming that there is no hardware noise.}
\subsubsection{Maximum Cut}

To create the plots in Figure~\ref{fig:MaxCut}, we prepared Erd\H{o}s-Rényi random graphs on $15$ nodes and generated their multigrid hierarchies.  Running each subgraph through Multigrid VQE and Efficient SU(2) VQE, we can see how the methods diverge in performance before reaching their solutions.  Some subgraphs had MaxCut zero (due to having no edges), and those datapoints were thrown away for Figure~\ref{fig:MaxCut_a} since their approximation ratios are undefined.

For MaxCut, we expect saving information about subgraph MaxCut to inform the optimization in subsequent refinement stages and improve performance.  Using NetworkX, we generated $15$-node Erdős-Rényi random graphs with edge probabilities $p\in\{0.3, 0.6, 0.9\}$ and compare approximation ratios between the Efficient SU(2)-seeded Multigrid VQE and the Efficient SU(2) VQE.

\subsubsection{Maximum $k$-Satisfiability}
To create the plots in Figure~\ref{fig:ksat}, we prepared  random E$k$-SAT instances at the satisfiability threshold for $15$ variables and generated their multigrid hierarchies.  Running each sub-formula through Multigrid VQE and Efficient SU(2) VQE, we can see how the methods diverge in performance before reaching their solutions.  It is worth noting that some sub-formulas had MaxSat zero (due to every clause sharing some not-yet-added variable) and that those datapoints were thrown away since their approximation ratios are undefined.  Further, there is no data for $2$ variables since there is no E$3$-SAT instance with only $2$ variables.

For Max E$k$-SAT, we expect the increased parameterization of the Multigrid Hierarchy and saving information about sub-formula MaxSat to inform the optimization in subsequent refinement stages and improve performance.  This appears to be achieved, as in Figure~\ref{fig:ksat} the multigrid methods outperform their hardware-efficient counterparts.

\section{Conclusion}
\label{sec:conclusion}
We have presented the hierarchical multi-grid ansatz as an alternative approach for the variational quantum eigensolver. Our approach generalizes classical mesh refinement, and can be applied anywhere a hierarchy of problems can be defined. Using a particular seed ansatz and multigrid refinement method, we showed improved average performance by comparison with the Efficient SU(2) VQE  for the Laplacian eigenvalue problem, as well as for MaxCut and Max-E$k$-Sat. 

\textcolor{black}{\textit{Future work:} }We will study this ansatz further for other application domains \textcolor{black}{such as quantum chemistry, where physical Hamiltonians such as the molecular electronic Hamiltonian could perhaps gain from exploiting the structure or symmetries of molecules in a multigrid hierarchy.}  Additional studies remain to be done to compare the multigrid ansatz for optimization problems to other quantum optimization algorithms\textcolor{black}{, such as QAOA variants and filtered VQE}. Also, in the future, we plan to complement our numerical studies with actual hardware studies on existing NISQ devices, \textcolor{black}{and to explore more efficient ansatz designs such as ansaetze tailored to qubit architectures without all-to-all connectivity.}

\appendix
\label{appendix}

In this appendix, we present the derivation of the Discrete Laplacian Hamiltonians.  To begin with, Poisson's equation $\nabla^2f=g$ can be discretized by second difference methods.  Assuming the one-dimensional case with unit displacement, we have the approximation
\[g_i=\nabla^2f_i=\partial f/\partial x\approx -f_{i-1}+2f_i-f_{i+1}.\]

\noindent In terms of vector entries, this translates to the formula
\[\sum_{i=0}^{n} (-f_{i-1} +2f_i-f_{i+1}){\bf e}_i\approx \nabla^2 f\approx \sum_{i=0}^{n} g_i{\bf e}_i,\]

\noindent where, taking Dirichlet boundary $f_{-1}=f_{n+1}=0$, the left-hand side motivates the discrete Dirichlet Laplacian.  This is the linear operator

\[\nabla^2_D=\begin{pmatrix}
2 & -1 &  &  &  &  &  \\
-1 & 2 & -1 &  &  &  &  \\
 & -1 & 2 & -1 &  &  &  \\
 &  & \ddots & \ddots & \ddots &  &  \\
 &  &  & -1 & 2 & -1 & \\
 &  &  &  & -1 & 2 & -1 \\
 &  &  &  &  & -1 & 2 
\end{pmatrix}\]

\noindent producing this matrix from the vector $f$.  Using periodic boundary $f_{-1}=f_{n-1}$, $f_{n}=f_0$, we instead get the discrete periodic Laplacian
\[\nabla^2_P=\begin{pmatrix}
2 & -1 &  &  &  &  & -1 \\
-1 & 2 & -1 &  &  &  &  \\
 & -1 & 2 & -1 &  &  &  \\
 &  & \ddots & \ddots & \ddots &  &  \\
 &  &  & -1 & 2 & -1 & \\
 &  &  &  & -1 & 2 & -1 \\
 -1 &  &  &  &  & -1 & 2 
\end{pmatrix}\]

\noindent which can be seen by matrix multiplication to satisfy the relationship $\nabla^2_D=\nabla^2_P+P^\dagger(I^{\otimes n-1}\otimes X)P$ from Sato et al.~\cite{sato2021variational}.  The same paper gives a third Laplacian for Neumann boundary conditions. Assuming zero derivative at the boundary, the second difference approximation changes to 
\[g_0=\nabla^2f_0=\partial f/\partial x\approx f_{-1}-f_{1}\]
\[g_{n}=\nabla^2f_{n}=\partial f/\partial x\approx f_{n}-f_{n-1}\]

\noindent so that the discrete Laplacian operator is given by
\[\nabla^2_N=\begin{pmatrix}
1 & -1 &  &  &  &  &  \\
-1 & 2 & -1 &  &  &  &  \\
 & -1 & 2 & -1 &  &  &  \\
 &  & \ddots & \ddots & \ddots &  &  \\
 &  &  & -1 & 2 & -1 & \\
 &  &  &  & -1 & 2 & -1 \\
  &  &  &  &  & -1 & 1 
\end{pmatrix}\]

\noindent completing the set of Laplacians from Sato et al.~\cite{sato2021variational}.  Future work could look at adapting these methods to more complicated boundary conditions.

\bibliographystyle{plainurl}
\bibliography{hierarchical-multigrid-ansatz-bib}

\end{document}